# Bulk Ferroelectric Heterostructures: Imprinted Actuators

Yizhe Li[1,2]*, Ziqi Yang[1,2], Ying Chen[1,2], Zhenbo Zhang[3], Yun-Long Tang[4], Annette K. Kleppe[5], Egor Koemets[5], Xuezhen Cao[1,2], Steven J. Milne[5], Juncheng Pan[1,2], Jiajun Shi[1,2], Yuge Yang[1,2], David A. Hall [1,2]*

[1]Department of Materials, University of Manchester, M13 9PL, Manchester, UK. [2]Henry Royce Institute, University of Manchester, M13 9PL, Manchester, UK. [3]Center for Adaptive System Engineering, School of Creativity and Arts, ShanghaiTech University, Shanghai, 201210, China. [4]Shenyang National Laboratory for Materials Science, Institute of Metal Research, Chinese Academy of Sciences, Wenhua Road 72, Shenyang 110016, China. [5]Diamond Light Source Ltd, Harwell Science and Innovation Campus, OX11 0DE, Didcot, UK [6]School of Chemical and Process Engineering, University of Leeds, LS2 9JT, Leeds, UK. *e-mail: yizhe.li@manchester.ac.uk; david.a.hall@manchester.ac.uk

**Abstract**

Domain switching is the cornerstone of ferroelectric materials. Most associated functionalities can be tuned via domain switching, including but not limited to piezoelectricity, electrostrain, thermal conductivity, electrocaloricity, domain wall conductivity and topological structures[1–5]. However, achieving the full potential of reversible ferroelectric domain switching is restricted by the incomplete access to the entire ferroelectric texture space, as well as the memory effects and energy dissipation associated with the hysteretic nature of ferroelectrics[6]. The manipulation of domain switching behaviour is moderately attainable in epitaxial heterostructures by exploiting the valence or lattice mismatch at heterointerfaces, which is generally constrained by the necessity for two dimensional architectures. In this study, domain-engineered bulk ferroelectric heterostructures (DE-BFH), constructed via elemental partitioning, are employed to unleash the full potential of bulk ferroelectrics, providing comprehensive control of domain switching characteristics and adjustable reversibility within the entire range of ferroelectric texture space. The exemplar DE-BFH ceramics exhibit unprecedented enhancement in reversible electrostrain and stability in both axial and shear modes, including a record high peak to peak shear strain up to 0.9% at intermediate field levels, confirmed by digital image correlation (DIC) measurements and in-situ synchrotron X-ray diffraction studies. The advancement of domain switching behaviour in DE-BFH could also promote the development of new types of lead-free piezoelectric devices, including actuators, energy harvesters, energy storage capacitors, multiple state memory devices, thermal switches, and domain wall switch. Moreover, the design concept of DE-BFH could contribute to the creation of distinctive ferroelastic, ferromagnetic, and multiferroic materials by broadening its scope to the entire ferroic family, encompassing polycrystalline, single-crystal, and thin-film forms.

The texture (domain orientation distribution, DOD) of a ferroelectric material can be rapidly and permanently reoriented by an applied electric field with magnitude close to or beyond the coercive field, which results in the signature hysteresis behaviour in ferroelectric materials (Fig. 1a). This characteristic feature determines that ferroelectric texture only acts as a transient attribute of ferroelectric materials in comparison with the persistent crystallographic texture established via texture engineering, such as templated grain growth or directional single crystal growth[7]. Hence, any electromechanical functionality, such as the direct or converse piezoelectric effects, established by ferroelectric domain engineering (poling) in ferroelectric materials, normally operates at the sub-coercive field ($E_c$) or stress ($\sigma_c$) condition to avoid irreversible erasure of the established ferroelectric texture along with the loss of functionality (Fig. 1b)[8]. These irreversible re-poling and depolarisation effects also induce significant drift of the origin point (variation in remnant strain) of piezoelectric actuators operating at field levels close to the coercive field (Fig. 1c)[9]. Additionally, axial actuators operating beyond the coercive field (in unipolar mode) achieve enhanced electrostrain via ferroelectric domain switching, albeit with partial ferroelectric texture variation (limited domain switching) accompanied by pronounced ferroelectric hysteresis as shown in Fig. 1a. These drawbacks significantly limit the output mechanical energy density, maximum strain level and motion accuracy of ferroelectric materials to fulfil their potential for electromechanical actuator applications[6]. The straightforward solution to these challenges could be the incorporation of highly reversible ferroelectric domain switching, which has the capability to recover the pre-set ferroelectric texture state following the removal of the external electric field. The same methodology can be extended to control additional functionalities governed by domain switching, such as thermal conductivity, domain-wall conductivity, and electro-optic properties.

The established methods for control of ferroelectric texture by the application of electrical and/or mechanical stresses, denoted here as electrical-domain engineering, E-DE, and mechanical-domain engineering, M-DE, are illustrated in Fig. 1e,f. In this respect, there is currently no feasible approach to effectively imprint or 'lock-in' the ferroelectric texture established in bulk materials via domain engineering, although partial domain stabilisation can be induced by the presence of oxygen vacancies or related defect-dipole complexes in so-called 'hard' ferroelectrics[10]. However the low energy defect nature (barrier) of such dipolar defect associates can still be overcome relatively easily under a high applied electric field, which induces inevitable ferroelectric 'deageing' along with the loss of the memorised ferroelectric texture; such effects are accelerated at moderately elevated temperatures[11,12]. In addition to introducing charged point defects, the exploitation of interfacial polarisation discontinuities and strain or compositional gradients within epitaxial heterostructures also exhibits the potential to lock-in the ferroelectric texture, which results in the shift of the polarisation-electric field (P-E) loops along electric field axis or constricted antiferroelectric-like hysteresis loops[13–15]. These mechanisms explicitly rely on the lattice and valence mismatch in delicately designed heterointerfaces, which are generally below 100 nm in thickness to avoid relaxation effects[15]. Therefore, it is challenging transfer such principles to bulk materials, in order to achieve similar functionalities in three dimensions rather than the out of plane direction of thin films.

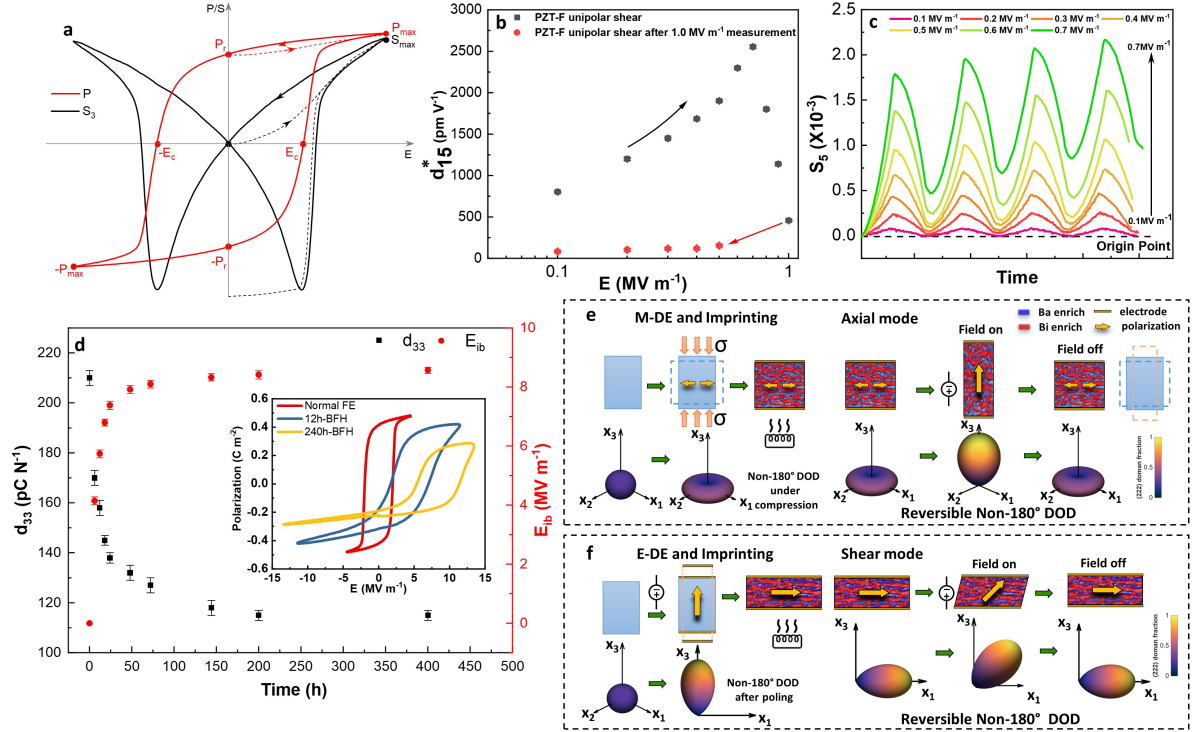

Fig. 1 **Limitations of normal ferroelectrics and creation of M-DE and E-DE BFH ceramics**. **a**, Illustration of polarisation and strain hysteresis behaviours in normal ferroelectrics under both unipolar and bipolar electric fields. **b**, effective shear piezoelectric coefficient, $d^*_{15}$, of soft PZT ceramics measured at various electric field levels up to 1 MV m$^{-1}$ and the repetition of the measurements after high field cycling. **c**, Variations in shear strain, $S_5$, of soft PZT ceramics under 4 consecutive cycles of unipolar field at different levels, showing significant shift in the origin point due to gradual re-poling at high field levels. **d**, Variations in internal bias field ($E_{ib}$) and piezoelectric coefficient ($d_{33}$) in prepoled BFH ceramics as a function of the duration of thermal ageing at 470 °C. **e** and **f**, Illustration of methods used to create mechanical-domain engineered (M-DE) and electrical-domain engineered (E-DE) BFH ceramics with imprinted domain orientation distribution (DOD) by exploiting the internal bias field induced via thermal ageing, as shown in (d). The 'locked-in' domain configuration permits reversible domain switching back to the imprinted DOD in both axial and shear driven modes under the electric field. The non-180° DOD surfaces presented in (e,f) were obtained by analysis of associated in-situ synchrotron x-ray diffraction experiments.

Our exploitation of the bottom-up synthesis approach for bulk ferroelectric heterostructures (BFH) via spontaneous elemental partitioning enables a highly tuneable and versatile routine to induce an internal bias field and imprint ferroelectric texture via straightforward heat treatment as demonstrated in Fig. 1e,f. The strength of the imprinting effect can be conveniently tuned by changing the thermal ageing time, as shown in Fig. 1d, yielding an internal bias field in poled samples of 0.7BiFeO$_3$-0.3BaTiO$_3$ ceramics up to 8.5 MV m$^{-1}$. The direction of internal field can be either random (symmetric constricted P-E loops) or unidirectional (asymmetric biased P-E loops), depending on whether a macroscopic polarity is established by the domain engineering process. In this study, we demonstrate the full control of domain switching by exploiting the combination of domain engineering and development of BFH driven by immiscibility. The exemplar axial and shear piezoelectric ceramics developed with the associated approaches exhibit unprecedented enhancement in electromechanical performance by utilising the full ferroelectric texture range and suppressing the limiting factor of hysteresis.

**Axial Mode Actuation**

Axial mode (global axes demonstrated in Fig 2.h) is the most common operation arrangement for ferroelectric materials. All the functionalities tuned via domain switching originate from either an unpoled (strain free and zero net polarisation) or a poled (tensile strain and remanent polarisation) state as shown in Fig. 1a, which reduces the available proportion of electric field induced domain switching (by at least 25%) and inhibits the full potential for control of functional performance by domain switching.

In this study, the attainable domain switching (towards $X_3$, $\psi=0°$) of $0.7BiFeO_3$-$0.3BaTiO_3$ (0.7BF-0.3BT) ceramics is enhanced by commencing from a remanent mechanically-depoled state (nearly theoretical minimum ferroelectric texture) achieved by means of M-DE. A new 'compressive' strain state (Fig. 1e and Fig. 2h) is established first via uniaxial compression of 0.7BF-0.3BT ceramics at elevated temperature (details in Methods) to attain saturated non-180° domain switching perpendicular to the $X_3$ direction and maximum remanent compressive strain up to -1.75 $\times 10^{-3}$, as shown in Extended Data Fig. 1a. The compressive strain state was subsequently stabilised (imprinted) by thermal ageing at 470 °C for 3 hours under a constant mechanical load, promoting elemental partitioning and BFH formation, as illustrated in Fig. 1e (details in Methods). This locked-in compressive strain state can be recovered and reset after the application of an electric field along $X_3$, $\psi=0°$, via spontaneous reverse domain switching that exploits the full ferroelectric texture space (i.e. the complete range of ferroelectric domain-switching capability), thereby enhancing the total attainable axial strain, $S_3$, as shown in Fig. 2h. The recovery of the remanent compressive state by reversible domain switching takes place upon removal of the electric field, facilitated by the local electric field and elastic stress imposed by aliovalent elemental partitioning within the BFH 0.7BF-0.3BT ceramic [16].

In order to verify the enhanced domain switching behaviour with improved reversibility, the 0.7BF-0.3BT ceramics in normal ferroelectric (FE), BFH and M-DE BHF states were investigated by in-situ synchrotron X-ray diffraction (XRD). The BFH state was simply created by thermal ageing the unpoled 0.7BF-0.3BT ceramics at 470 °C for 3 hours without any pre-domain engineering. All of the 0.7BF-0.3BT ceramics with different states were found to exhibit a rhombohedral structure at the long-range macroscopic scale with characteristic double peak splitting of {222} reflections (Fig. 2e-g). The evolution of the intensities of (222) and ($\bar{2}22$) reflections associated with the respective polar and non-polar directions exhibit evident 'digital-like'[6] switching behaviour with field on and off in both BFH and M-DE BFH states, whereas the intensity variation of {222} reflections in normal FE 0.7BF-0.3BT significantly lags behind the electric field waveform showing strong hysteresis as shown in Fig. 2 a-c & e-g.

The domain switching behaviour is further quantified in terms of the domain fraction, $\eta_{hkl}(\psi)$, in line with the in-situ XRD results (details in Methods). In the unpoled normal FE state, the maximum attainable domain switching fraction along the axial direction for the polar pole, [222], of a rhombohedral phase is $\Delta\eta_{222}(\psi=0°) = 0.75$ (Fig. 2j), which is about $\Delta\eta_{002}(\psi=0°) = 0.67$ for the polar pole, [002], for a tetragonal phase[17]. However, the achievement of this maximum value is hindered by the constraint of enhanced intergranular stress under high strain levels[18,19]. As a result, a saturated domain switching fraction of $\Delta\eta_{222}(\psi=0°) = 0.7$ was achieved in the normal FE state 0.7BF-0.3BT (Fig. 2j), and the fraction of reversible domain switching under unipolar field is only about 0.07. In comparison, the domain switching behaviour is significantly improved in the M-DE BHF state, exhibiting a 26% enhancement in $\Delta\eta_{222}(\psi=0°)$ up to 0.88, due to the 'locked-in' compressive strain of the mechanically-depoled state (Fig. 2j). The reversible domain switching fraction under unipolar field is also increased to 0.77, which is about one order of magnitude higher than that of the normal FE state. As a result, a 'digital-like' S-E response is observed for both axial and transverse electrostrains of M-DE BHF material as shown by the variations in the macroscopic strain orientation distribution (SOD) in Fig. 2d.

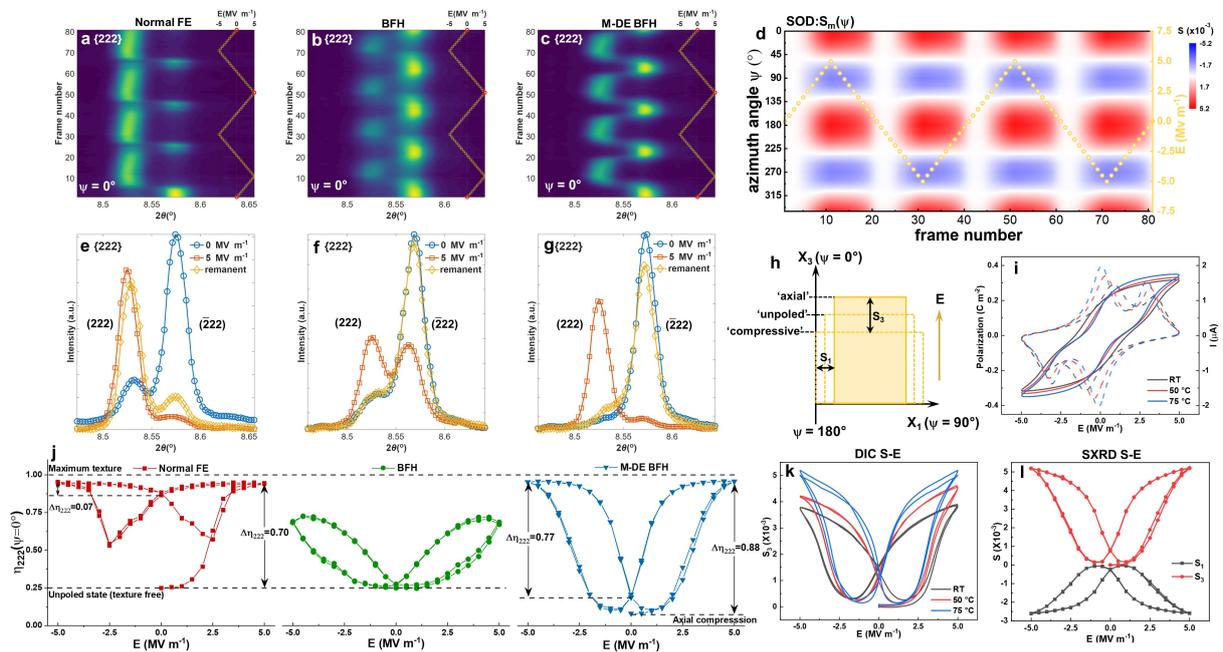

Fig. 2 **Axial electromechanical behaviour of mechanical-domain engineered bulk ferroelectric heterostructure (M-DE BFH) ceramics**. **a-c**, the contour plots of X-ray diffraction (XRD) peaks for the evolution of {222} reflections in the axial direction ($\psi=0°$) under the bipolar electric field for normal ferroelectric (FE), BFH and M-DE BFH ceramics at 75 °C. **d,** the evolution of azimuth angle dependent macroscopic strain distribution (SOD), $S_m(\psi)$, of M-DE BFH ceramics with the applied electric field at 75 °C. The evaluation of $S_m(\psi)$ is based on the associated synchrotron results. **e-g**, the in-situ X-ray diffraction profiles of {222} reflections at the initial state without field (0 MV m$^{-1}$), for an electric field of 5 MV m$^{-1}$, and in remanent state for normal FE, BFH and M-DE BFH ceramics respectively, selected from the results from **(a)-(c)**, and highlighted by red circles in the profile of electric field waveforms. **h**, illustration of the electrostrain behaviour of M-DE BFH and the definition of global axes. **i**, polarisation- and current-electric field (P-E and I-E) loops of M-DE BFH ceramics at room temperature, 50 and 75 °C. **j**, electric field-dependent domain fraction of (222) poles, $\eta_{222}$, for normal FE, BHF and M-DE BFH ceramics. **k**, Strain-electric field (S-E) loops of M-DE BFH ceramics measured by digital image correlation (DIC) method at room temperature, 50 and 75 °C respectively. **l**, the S-E loops of M-DE BFH ceramics evaluated according to in-situ synchrotron X-ray diffraction data obtained at 75 °C.

Similarly, constricted P-E loops were observed over a range of temperatures in the M-DE BFH ceramics, reflecting enhanced recovery of domain switching (Fig. 2i), in contrast to the rectangular P-E loop of the normal FE BF-BT ceramics (Extended Data Fig. 1b). Further digital image correlation (DIC) measurements, conducted under the same condition as the in-situ XRD experiments at 75°C, reveal that the associated total achievable axial electrostrain, $S_3$ (up to 5.2 x$10^{-3}$, Fig. 2k), of M-DE BFH is significantly enhanced with the recovery of the remanent compressive strain (1.75 x$10^{-3}$) in addition to the total strain for the normal FE BF-BT material (3.5 x$10^{-3}$), as shown in Extended Data Fig. 1a&b. The consistency of reversibility and magnitude of electrostrain for M-DE BFH ceramics determined by both DIC measurements and in-situ XRD estimation (Fig. 2k&l) confirmed the enhancement in tuning electrostrain behaviour by utilizing the complete range of ferroelectric texture space. The DE BFH design enables precise control over domain-switching functionalities via tailored ferroelectric textures, tuning properties such as polarisation, thermal conductivity, and electrical conductivity. The combination of macroscopic DIC strain measurements together with evaluation of the contributing mechanisms by in-situ XRD offers a robust strain-measurement framework that eliminates any possible misinterpretation of apparent strain induced by bending displacements[20], which may result from uneven electric-field distributions in semiconductive ferroelectrics[21].

**Shear Mode**

Shear mode piezoelectric devices outperform their axial mode counterparts in many key piezoelectric parameters. However, the level of output shear strain is limited due to the presence of depoling or re-poling effects close to or beyond the coercive field, which impedes the application of shear mode piezoelectrics as high displacement actuators or the miniaturisation of associated devices. The exploitation of cross-poled BFH 0.7BF-0.3BT ceramics via E-DE and elemental partitioning (Fig. 1f) could facilitate the creation and manipulation of reversible domain switching beyond axial direction, and which also unlocks the output limit on shear strain of ferroelectrics.

For operation in the shear mode, a moderate internal bias field, $E_{ib}$=1.5 MV m$^{-1}$, was induced along the transverse direction ($X_1$) of and E-DE BFH ceramics by annealing at 470 °C for 3 h (global axes demonstrated in Fig 3.a). This resulted in a shifted asymmetric transverse $P_1$-E loop and a constricted symmetric axial $P_3$-E loop (Fig. 3b), indicating the occurrence of reversible domain switching from $X_3$ direction to $X_1$ direction upon field removal. The resulting shear strains ($S_5$, Fig. 3c,d) , measured by DIC, reached 4.25 x$10^{-3}$ (unipolar) and 8.5 x$10^{-3}$ (bipolar, peak to peak) for an applied field of 4.5 MV m$^{-1}$, exceeding the maximum shear strains ($S_5$) achieved in conventional soft or hard PZT by more than fourfold[8]. Such shear mode operation is generally inaccessible in thin-film devices, which are limited to purely in-plane or out-of-plane axial actuation modes. The attainable shear strain in E-DE BFH ceramics is also maximised by the recovery of domain switching from the 'field-on' state ($\psi$ = 45°) back to the $X_1$ direction upon removal of the applied electric field, overcoming the limitation of memory effects from re-poling or cross poling and yielding pinched $S_5$-E loops (Fig. 3c) with 'digital-like' displacement behaviour. The distorted tip of the $S_5$-E loop measured at 5 MV m$^{-1}$ is attributed to the rotation of the DOD and shift of principle strain axis away from $\psi$=45° towards the $X_3$ direction (Fig. 3c,g).

For BiFeO$_3$-BaTiO$_3$ solid solutions, it was observed that the aliovalent elemental partitioning process is most prominent at elevated temperatures between 440 and 800 °C. Below this temperature range, the A-site cations (Bi$^{3+}$ and Ba$^{2+}$ in this case) are essentially immobile. Consequently, the BFH 0.7BF-0.3BT ceramics do not suffer from the electric field-driven 'deageing' effects that are commonly observed in acceptor-doped hard ferroelectrics, in which the stabilisation of domain walls is achieved by the reorientation of relatively mobile acceptor ion-oxygen vacancy defect associates[11,12]. Therefore, the electric field induced shear strain in BFH 0.7BF-0.3BT ceramics exhibits excellent stability under either unipolar or bipolar field cycling, and shows no sign of deageing at either room or elevated temperatures as shown in Fig. 3 d and Extended Data Fig. 2a,b. The shear mode electromechanical performance is further enhanced with increasing temperature; the highest effective converse shear piezoelectric coefficient, $d_{15}^*$, increases from about 1000 to 1370 pm V$^{-1}$ along with the peak to peak shear strain from 8.5 x10$^{-3}$ to 9.8 x10$^{-3}$ at RT and 100 °C respectively (Fig. 3e and Extended Data Fig. 2c). The reversible domain switching also ensures the retention of the low field shear piezoelectricity after exposing to a field above $E_c$ (Extended Data Fig. 2d), in comparison with the loss of shear strain in normal ferroelectrics after cross-poling (Fig 1.b). The reversibility of shear strain can be further tuned by modifying the level of internal bias field by the selection of different thermal ageing time (Extended Data Fig. 2e&f).

Further in-situ synchrotron x-ray diffraction measurements (SXRD) were performed to validate the DIC strain results and reveal the micromechanical origin of the electric field-induced shear strain in E-DE BFH 0.7BF-0.3BT ceramics. In this work, an azimuthal angle of ψ=45° was selected to indicate the development of lattice shear strain by the shift of the singular {200} peak, while variations in the relative intensities of the (222) and ($\bar{2}$22) reflections, corresponding to the polar and non-polar directions respectively, provide the means to quantify the degree of non-180° domain switching (Fig. 3f). Significant domain switching behaviour was observed only for electric field levels above 2.5 MV m$^{-1}$. Below $E_c$, the electromechanical strain is dominated by the intrinsic piezoelectric effect, giving rise to lattice strain with negligible domain switching (Fig. 3g,j). The complete reversibility of non-180° domain switching is confirmed by the almost identical peak profiles of {222} reflections before and after the exposure to an electric field 5 MV m$^{-1}$, which is significantly greater than $E_c$ (Fig. 3i).

Although the orientation of principal strain, ψ$_p$, deviates from the shear direction (ψ=45°) above $E_c$ (Fig. 3g), the principal strain still significantly contributes to the improved shear strain tensor component, $S_{31}$, via enhanced domain switching along ψ=45° as shown in the domain orientation distribution (DOD) for the polar (222) direction in Fig. 3j (details for quantification of SOD, DOD and principal strain in Methods). The estimated field-induced shear strain $S_5$ ($S_5=2S_{31}$, Fig. 3a,c), extracted from the SOD (ψ=45°), agrees well with DIC results across the range of electric field levels (Fig. 4h), confirming the reliability of both DIC and SXRD measurements.

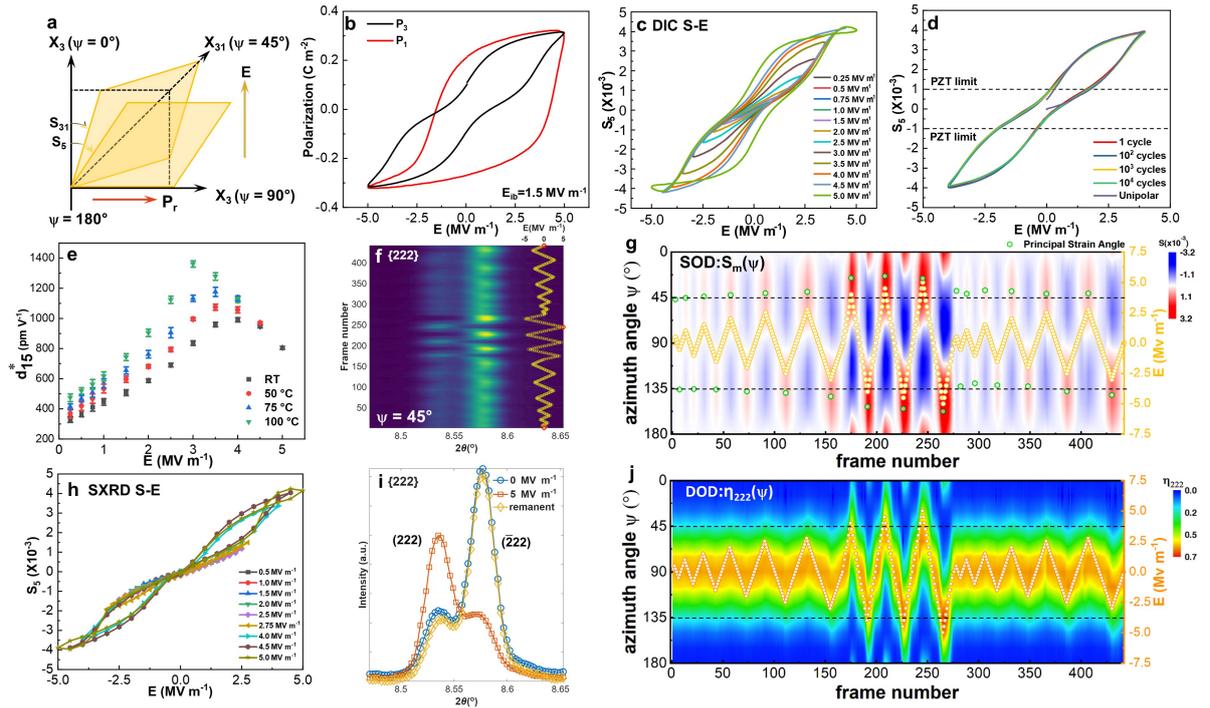

Fig. 3 **Shear electromechanical behaviour of electrical-domain engineered bulk ferroelectric heterostructures (E-DE BFH) ceramics**. **a**, illustration of the relationship between engineering shear strain, $S_5$, and tensorial shear strain, $S_{31}$ with respect to the definition of global axis and remanent polarisation direction. **b**, the polarisation-electric field (P-E) loops measured from axial ($P_3$-E) and transverse ($P_1$-E) directions respectively. **c**, shear strain-electric field loops ($S_5$-E) measured under different field levels up to 5 MV m$^{-1}$, in comparison with the $S_5$-E loops measured by SXRD in **h**. **d**, repetition of the $S_5$-E measurements at 4 MV m$^{-1}$ after the samples were cycled under 5 MV m$^{-1}$ for 1, $10^2$, $10^2$, $10^3$ and $10^4$ times, and an additional unipolar $S_5$-E measurement after the investigation on cycling stability. **e**, electric field dependent effective shear piezoelectric coefficient, $d_{15}^*$, of E-DE BFH ceramics measured at room temperature up to 100 °C. **f**, contour plot for the evolution of {222} reflections for E-DE BFH ceramics at $\psi=45°$ under a bipolar electric field waveform with progressively increasing magnitude The selected XRD profiles in the initial state without field (0 MV m$^{-1}$), at an electric field of 5 MV m$^{-1}$, and in the remanent state are presented in **i**. **g** and **j**, the evolution of azimuth angle dependent strain distribution (SOD) and domain orientation distribution (DOD) for the respective macroscopic strain, $S_m(\psi)$, and domain orientation fraction $\eta_{222}(\psi)$ under a bipolar electric field waveform with progressively increasing amplitude.

In summary, we bridge the gap between ferroelectric epitaxial heterostructures and bulk ferroelectrics by creating DE-BFH. The construction process of DE-BFH conveniently exploits the advantages of both full spatial domain engineering for bulk ferroelectrics and valence mismatch at epitaxial heterointerfaces by nanoscale elemental partitioning. Such unique characteristics of DE-BFH facilitates the creation and manipulation of reversible domain switching, utilising the entire range of ferroelectric texture configurations, as demonstrated by the exceptionally enhanced axial and shear electrostrains. In particular, DE-BFH BF-BT ceramics are effective lead-free alternatives for commercial PZT-based materials with enhanced strain output and extended application temperature range. The innovation of DE-BFH also has wide-ranging implications for improving control and performance of functionalities associated with reversible and programmable ferroelectric domain switching.

Furthermore, the design of similar bulk heterostructures for other types of ferroic materials could contribute to the development of unique ferroelastic, ferromagnetic, electro optical properties and the coupling of theses functionalities in monolithic bulk components.

**Methods**

**M-DE BFH sample preparation.** The uniaxial compression and thermal ageing experiments on normal FE $0.7BiFeO_3$-$0.3BaTiO_3$ (BF-BT) ceramics were performed with Instron 6800 Series Universal Testing System. The methods used for the preparation of the normal FE $0.7BiFeO_3$-$0.3BaTiO_3$ ceramics were described in our previous publications[16]. The samples were grind and polished to achieve parallel surfaces (accuracy <3 μm) for thermo-mechanical experiments. A square-ended beam-shaped sample (10x5x5 mm in dimensions) was carefully centred and aligned with the stress axis with the assistance of a centering tool at room temperature. The sample was then heated to 350 °C with a heating rate of 10 °C min$^{-1}$ to apply a constant compressive stress of 50 MPa. After dwelling for 10 min at 350 °C under 50 MPa compressive stress, the temperature was increased to 475 °C with a heating rate of 10 °C min$^{-1}$, retaining the constant compressive stress of 50 MPa. The sample was then thermally aged at 475 °C under 50 MPa compressive stress for 3 hours, followed by a cooling process with a cooling rate of 10 °C min$^{-1}$. The M-DE BFH Sample achieved via this thermomechanical treatment was further machined to 5x1x1 mm bars for further in-situ synchrotron and DIC experiments.

**E-DE BFH sample preparation.** The normal FE BF-BT ceramics were first machined to 5x1x1 mm bar shaped samples, followed by a poling step along the length direction. The poling procedures were performed at 80 °C with a 5 MV m$^{-1}$ electric field for 10 minutes. The poled samples were then re-electroded on the parallel rectangular surfaces using a silver paste (GWENT C2130823D1) as shown in Fig. 1f. Thermal ageing of the poled samples was performed at 475 °C for 3 and 4 hours with the same heating and cooling rate of 10 °C min$^{-1}$ to achieve internal bias fields about 1.5 and 1.75 MV m$^{-1}$ respectively.

**In-situ synchrotron XRD measurements.** In-situ synchrotron X-ray diffraction (SXRD) experiments were performed at Beamline I15, Diamond Light Source, UK with a photon energy of 72 keV, a beam diameter of 76 x 115 μm (vertical x horizontal) and using transmission geometry. The direction of applied Electric field (ψ=0°) is parallel with the axial direction ($X_3$, shown in Fig. 2h) of the M-DE BFH sample and transverse direction ($X_1$, shown in Fig. 3a) of the E-DE BFH sample respectively. 2D diffraction patterns were caked into 24 banks (15°/bank) along azimuthal angles, ψ from 0 to 360° and integrated into 24 corresponding 1-D line profiles. Further details of the experimental setup, data collection and post processing are provided in our previous study[1].

The Daymond method[1] was adopted to estimate the macroscopic strain at a given azimuthal angle, $S_m(\psi)$, taking account of the effective lattice strains for the representative grain orientations {222}, {200} and {220}, given as:

$$S_m(\psi) = \frac{\sum m_{\{hkl\}} S^*_{\{hkl\}}(\psi)}{\sum m_{\{hkl\}}} \quad (1)$$

where $m_{\{hkl\}}$ is the total multiplicity of grain family {hkl} for {222}, {200} and {220} respectively. The effective lattice strain for each grain family $S^*_{\{hkl\}}(\psi)$ is determined by the effective d-spacing $d^*_{\{hkl\}}(\psi)$, and $d^*_{\{hkl\}}(\psi)$ is obtained by a weighted average process, taking account of the domain fraction $\eta_{(hkl)}(\psi)$ and d-spacing of the reflection (hkl), $d_{(hkl)}(\psi)$, which can be respectively expressed as:

$$S^*_{\{hkl\}}(\psi) = \frac{d^*_{\{hkl\}}(\psi) - d^0_{\{hkl\}}(\psi)}{d^0_{\{hkl\}}(\psi)} \quad (2)$$

$$d^*_{\{hkl\}}(\psi) = \sum \eta_{(hkl)}(\psi) d_{(hkl)}(\psi) \quad (3)$$

where $d^0_{\{hkl\}}(\psi)$ is the initial effective lattice spacing at a given azimuthal angle, which acts as the relative zero strain lattice spacing at each bank in comparison with the strain free lattice spacing prior to domain engineering. The analysis methods used for detailed evaluation of the domain fractions of different grain families, $\eta_{(hkl)}(\psi)$, are given in our previous study[1]. The estimated macroscopic axial and transverse strains for M-DE BFH ceramics are achieved at ψ=0 and 90° for $S_m(\psi)$. The estimated macroscopic engineering shear strain, $S_5$, for E-DE BFH ceramics is double the magnitude of the tensorial shear strain, $S_{31}$, which is achieved at ψ=45° for $S_m(\psi)$. By considering the 90° rotation of the global axes and the deviation of principle strain away from the global axes in shear mode, the orientation distribution of macroscopic strain, $S_m(\psi)$, in E-DE BFH ceramics follows a modified second order spherical harmonic function[1], given as:

$$S_m(\psi) = (d'_{33} - d'_{31}) E \cos^2(\psi - \Delta\psi) + d'_{31} E \quad (4)$$

Where Δψ is the principal strain angle, $d'_{31}$ and $d'_{33}$ are the effective piezoelectric coefficients parallel and perpendicular to the principal strain direction respectively, and E is the strength of electric field. Therefore, the principal strain angle Δψ can be determined by nonlinear fitting of the experimental data of $S_m(\psi)$ with equation (4). The associated XRD peak profile fitting and the evaluation of DOD and SOD were performed with in-house developed MATLAB scripts.

**Digital Image Correlation (DIC) strain measurements.** The macroscopic strain-electric field (S-E) loops of E-DE and M-DE BFH samples were measured with a 2D DIC method under an optical microscope. A square-ended bar specimen was placed in a custom-designed sample holder with electrical contact to the HV amplifier. The images for strain calculation were acquired from the cross-section of the specimen immersed in silicone oil in order to avoid electrical arcing in air. A triangular electric field waveform with a frequency of 0.1 Hz was applied for all the measurements for compatibility with the image acquisition rate of 8 fps. The image acquisition, electric field application and recording were controlled by a LabVIEW program. The sequence of images was processed with commercial DIC software (LaVision Davis 10.2.1) to determine the 2-D deformation relative to the initial zero-field reference image. Further details of the DIC experiment setup and strain calculation can be found in our previous study[17].

**Extended Data**

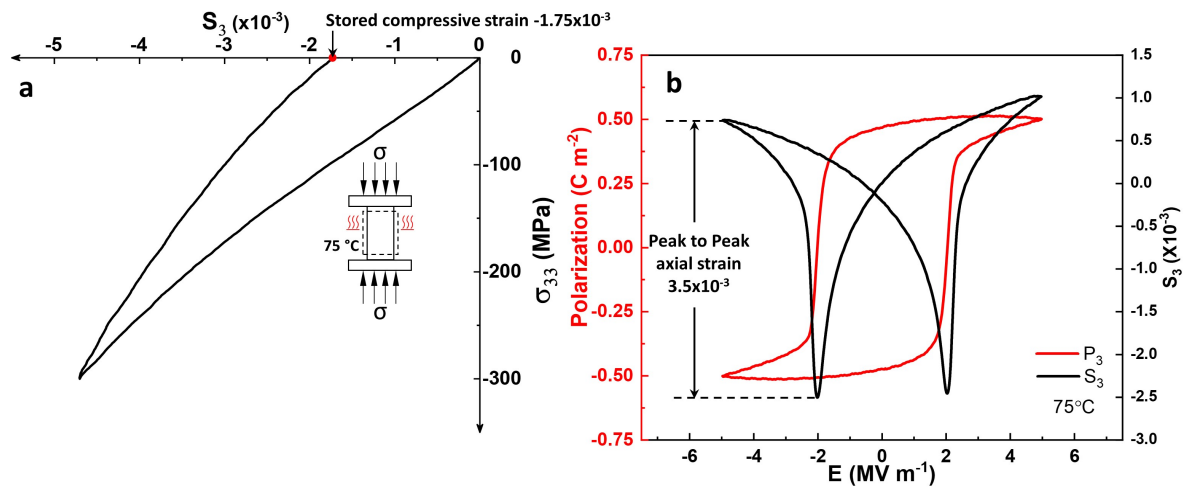

Extended Data Fig. 1 Strain response of 0.7BF-0.3BT under **a.** mechanical load and **b.** electrical field at 75 °C

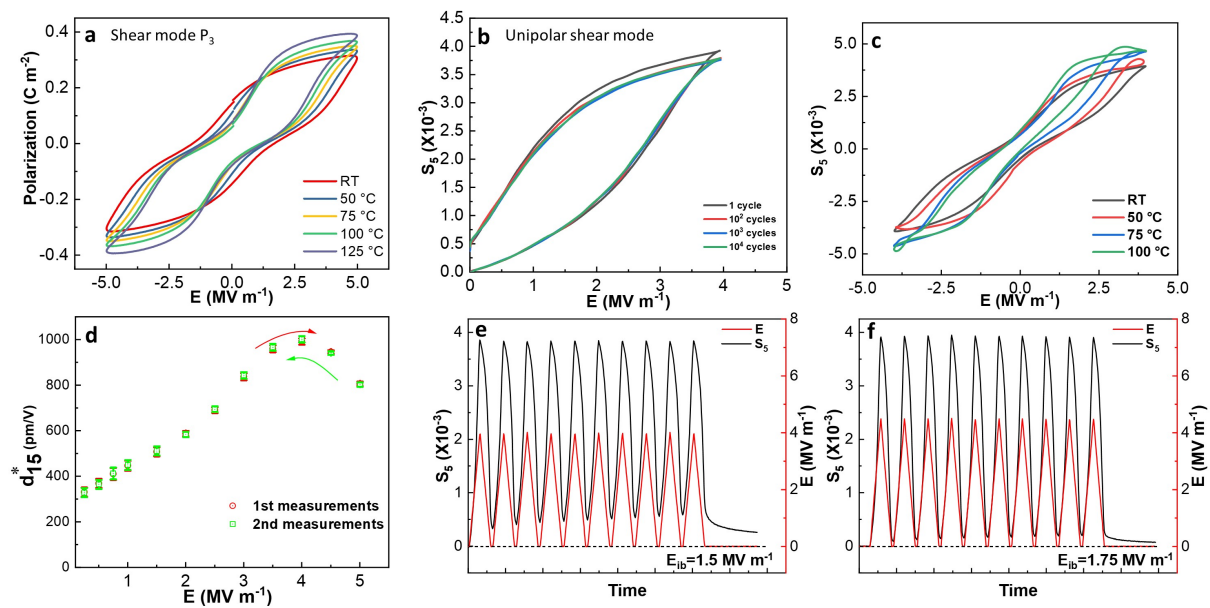

Extended Data Fig. 2 **Thermal stability and tunability of shear mode E-DE BFH. a.** temperature dependent axial $P_3$-E loops at 5 MV m$^{-1}$ measured from RT to 125 °C. **b.** repetition of the unipolar $S_5$-E measurements at 4 MV m$^{-1}$ after the samples cycled under 5 MV m$^{-1}$ for 1, $10^2$, $10^2$, $10^3$ and $10^4$ times. **C.** temperature dependent $S_5$-E loops at 3.75 MV m$^{-1}$ measured from RT to 100 °C. **d.** effective shear piezoelectric coefficient, $d_{15}^*$, of E-DE BFH ceramics measured at various electric field level up to 5 MV m$^{-1}$ and the repetition of the measurements after the exposure of 5 MV m$^{-1}$. **e** and **f**, the unipolar $S_5$ responses of E-DE BFH ceramics with $E_{ib}$=1.5 and 1.75 MV m$^{-1}$ respectively.